\documentclass[12pt]{article}
\usepackage{amsmath}
 \usepackage{paralist}
  \usepackage{graphics}
   \usepackage{epsfig}
     \usepackage{graphicx,subfigure}

\newcommand{\half}{\mbox{$\frac{1}{2}$}}

\begin{document}

\begin{center}
{\bf \Large Network Inference with Hidden Units}

\vspace*{0.5cm}
{\large Joanna Tyrcha\\}
{\small Mathematical Statistics, Stockholm University, S106 91 Stockholm, Sweden\\
{\tt joanna@math.su.se}}

\vspace*{0.2cm}
{\large John Hertz  \\}
{\small Niels Bohr Institute, Blegdamsvej 17, 2100 Copenhagen {\O}, Denmark\\
NORDITA, Roslagstullsbacken 23, 10691 Stockholm, Sweden\\
{\tt hertz@nordita.dk}}
\end{center}

 \vspace*{0.5cm}
\subsection*{Abstract}
We derive learning rules for finding the connections between units in stochastic dynamical networks from the recorded history of a ``visible'' subset of the units.  We consider two models. In both of them, the visible units are binary and stochastic.  In one model the ``hidden'' units are continuous-valued, with sigmoidal activation functions, and in the other they are binary and stochastic like the visible ones.  We derive exact learning rules for both cases.  For the stochastic case, performing the exact calculation requires, in general, repeated summations over an number of configurations that grows exponentially with the size of the system and the data length, which is not feasible for large systems.  We derive a mean field theory, based on a factorized ansatz for the distribution of hidden-unit states, which offers an attractive alternative for large systems.  We present the results of some numerical calculations that illustrate key features of the two models and, for the stochastic case, the exact and approximate calculations.

\newpage

\section{Introduction}

Recent interest in network identification problems has been motivated by the advent of multi-electrode neural recordings and other large-scale biological data \cite{Schneidman,RTHPRE09,RHPRL11,chapter}.  Current  inference methods, however, do not take into account the effects of units in the networks that are not recorded, though they are almost always present.  This problem can be serious:  For example,  in cortical neural data, almost all recorded cells are excitatory, though inhibitory cells are essential in the network dynamics.  In this paper we extend previous methodology to include ``hidden units'', presenting algorithms for inferring the strengths of connections to, from and among them.  

There is a long history of work of problems of this sort.  Perhaps the best know is that on ``Boltzmann machines'' \cite{AHS85}.   These are symmetrically coupled networks of stochastic binary units.  Their states are updated, one randomly chosen unit at a time, with the probability of being in a particular one of its two possible states given by a logistic sigmoid function of the net input from other units.  Because of the symmetric coupling matrix, their dynamics satisfies detailed balance, so their equilibrium distributions are of Gibbs-Boltzmann form $Z^{-1} \exp (-E)$, where $E$ is a quadratic form.  This fact that simplifies their analysis considerably.   The problem has also been studied in networks where the unit outputs are continuous sigmoidal functions of their inputs, for both continuous-time (asynchronous-update) and discrete-time (simultaneous-update) dynamics, extending the back-propagation algorithm used earlier for layered networks.  

Applying either of these kinds of models to multineuron spike data is problematic.   Real biological networks do not have symmetric connections, invalidating the first kind, while the nature of synaptic transmission and neuronal spiking calls for a stochastic binary representation, ruling out the second.  In this paper we treat models in which the recorded neurons are stochastic and binary, and there is no symmetry requirement on the  connections in the network.  They obey a discrete-time kinetic Ising (Glauber) dynamics \cite{Glauber}, and a value $+1$ represents an action potential.   We study two kinds of models, in which the hidden units are deterministic or stochastic, respectively.   We employ, for convenience, a discrete-time dynamics \cite{Peretto}, though it should be straightforward to extend the treatment to continuous-time models.

\section{Continuous, deterministic hidden units}

We examine first the deterministic case, taking the output of a hidden unit to be a sigmoidal function of its input.  Though it is a big simplification of a real spiking-neuron network, this kind of model can be practical for analyzing neural data.  One cannot hope to model the detailed dynamics of all the unrecorded neurons in the network of interest, because they vastly outnumber the recorded ones. What one can hope to do, at least as a first approximation, is to describe the effect of unrecorded \emph{populations} of neurons, for example, of inhibitory neurons when only excitatory neurons have been recorded.  The values of the hidden units in our model here represent the firing rates of those populations.  While this representation throws out many details, it enables one to capture some essential features of the dynamics, even using only one or a few hidden units.

We draw here on work in learning in analog neural networks a couple decades ago, under the names ``back-propagation in time'' and ``recurrent back-propagation'' \cite{RHW,Pineda,Pearlmutter,WilliamsZipser}.  Our treatment differs from that work in having stochastic visible units and a likelihood-based objective function.

\subsection{Model}

\begin{figure}[htp]
\begin{center}
 \includegraphics[width=5in]{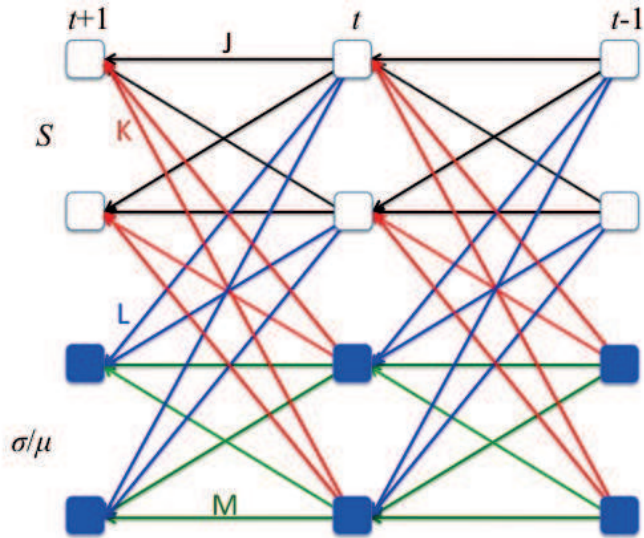}
  \caption{Schematic picture of the model.  (Color online) White squares represent visible units $s_i$; blue ones, hidden units $\mu_a$ (or $\sigma_a$ when they are stochastic).  Visible-visible connections $J_{ij}$ are black, hidden-to-visible ones $K_{ib}$ are red, visible-to-hidden ones $L_{aj}$ are blue, and hidden to hidden ones $M_{ab}$ are green.  Rows represent time steps.}     
 \label{netgraph}
 \end{center}
\end{figure}

We denote the states of the visible units by $s_i(t)$, where $i$ labels the unit and $t$ the time bin.   They can take the values $\pm 1$.  (We assume the recorded spikes have been sorted into time bins small enough that there is no more than one spike per bin.)  We denote the hidden unit values by $\mu_a(t)$, $\-1 \le \mu_a(t) \le 1$.  To make our equations a little more transparent, we use indices $i$, $j, \cdots$ for visible units and $a$, $b, \cdots$ for hidden ones.  Our model is defined by the stochastic evolution rule
\begin{eqnarray}
P[s_i(t+1)| \{s(t),\mu(t)\}] &=& \frac{\exp [s_i(t+1)H_i(t)]}{2 \cosh H_i(t)}			\label{visdyn} \\
\mu_a(t+1) &=& \tanh B_a(t),											\label{hiddyn}
\end{eqnarray}
with
\begin{eqnarray}
H_i(t) &=& \sum_j J_{ij}s_j(t) + \sum_b K_{ib}\mu_b(t)						\label{Hdef} \\
B_a(t) &= &\sum_j L_{aj}s_j(t) + \sum_b M_{ab} \mu_b(t).						\label{Bdef}
\end{eqnarray}
All $s_i(t+1)$ are assumed independent, conditional on $\{s_j(t)\},\{\mu_b(t)\}$.  The model is pictured in Fig.~\ref{netgraph}.  We do not write constant bias terms in $H$ or $B$ here; they can be included by adding input units which are always $+1$.  We will denote the number of visible units by $N_v$ and the number of hidden ones by $N_h$.

\subsection{Objective function and learning rules}

We assume we are given the data $\{s_i(t)\}$ and that we know the number of hidden units.  The task is to learn the connections $\{J_{ij}\}$, $\{K_{ia}\}$, $\{L_{aj}\}$, and $\{M_{ab}\}$, and our objective function is the log likelihood of the observed visible history:
\begin{equation}
{\mathcal L} = \sum_{it} [s_i(t+1)H_i(t) - \log 2 \cosh H_i(t)].					\label{LL}
\end{equation}

We consider the simplest form of gradient-based learning, where the parameters are adjusted proportional to the derivative of the log likelihood with respect to them.  For $\{J_{ij}\}$ and $\{K_{ia}\}$, this is straightforward:
\begin{eqnarray}
\Delta J_{kl} &=& \sum_{it}[s_i(t+1) - \tanh(H_i(t))] 
\frac{\partial H_i(t)}{\partial J_{kl}}  
= \sum_{t} \epsilon_k(t+1) s_l(t),						\label{learnJ} \\
\Delta K_{kb} &=& \sum_{it}[s_i(t+1) - \tanh(H_i(t))] 
\frac{\partial H_i(t)}{\partial K_{kb}} 							 
= \sum_{t} \epsilon_k(t+1) \mu_b(t),						\label{learnK}
\end{eqnarray}
with $\epsilon_k(t+1) = s_i(t) - \tanh H_i(t)$, the observed error on unit $i$ at $t+1$ under the model with the current parameters, given its state at $t$.  This is standard error $\times$ input learning, as in networks without hidden units \cite{RHW,RHPRL11}.

For the connections  that lead to hidden units, the derivatives of $H_i(t)$ with respect to $\{L_{aj}\}$ and $\{M_{ab}\}$ are through its dependence on the $\mu_b(t)$; as in
\begin{equation}
\Delta L_{al} = \sum_{it} \epsilon_i(t+1)\frac{\partial H_i(t)}{\partial L_{al}} 
= \sum_{it} \epsilon_i(t+1) \sum_b K_{ib} \frac{\partial \mu_b(t)}{\partial L_{al}}.		\label{learnL}
\end{equation}
Furthermore, the derivatives of the $\mu_j(t)$ have terms proportional to derivatives of all the $\mu$s at the previous time step:
\begin{equation}
 \frac{\partial \mu_b(t)}{\partial L_{al}} =(1-\mu_b^2(t))\left[  \delta_{ab} s_l(t-1)
+ \sum_c M_{bc} \frac{\partial \mu_c(t-1)}{\partial L_{al}} \right].				\label{dmdLrecrel}
\end{equation}

These equations can be iterated starting from the initial condition $\partial \mu_c(0)/\partial L_{al} = 0$.  The solution can be written relatively compactly:
\begin{equation}
 \frac{\partial \mu_b(t)}{\partial L_{al}}  = X_{bb}(t)\left \{\delta_{ab}s_l(t-1) 
+ \sum_{q=1}^{t-1} \left[ \prod_{r=1}^q [{\sf MX}(t-r)]\right]_{ba}s_l(t-q-1) \right\},  \label{gendmdL}
\end{equation}
where
\begin{equation}
X_{ab}(t) = (1-\mu_a^2(t))\delta_{ab}									\label{defX}
\end{equation}
and we make the convention that the product over $r$ is equal to $1$ when $q=0$.  The learning rule for $L_{al}$ can then be written as
\begin{equation}
\Delta L_{al} = \sum_{t}
 \sum_{q=0}^{t-1}\sum_i\epsilon_i(t+2+q)\left [{\sf KX}(t+1+q) \left( \prod_{r=1}^q {\sf MX}(t+r)\right) \right]_{ia}s_l(t), 
 															\label{dLresult}
\end{equation}
Exactly the same procedure for the derivative with respect to $M_{ab}$ gives
\begin{equation}
\Delta M_{ab} = \sum_{t}
 \sum_{q=0}^{t-1}\sum_i\epsilon_i(t+2+q)\left [{\sf KX}(t+1+q) \left( \prod_{r=1}^q {\sf MX}(t+r)\right) \right]_{ia}\mu_b(t),  
 															\label{dMresult}
\end{equation}
which differs from (\ref{dLresult}) only in the last factor.

\begin{figure}[htp]
\begin{center}
\includegraphics[width=5in]{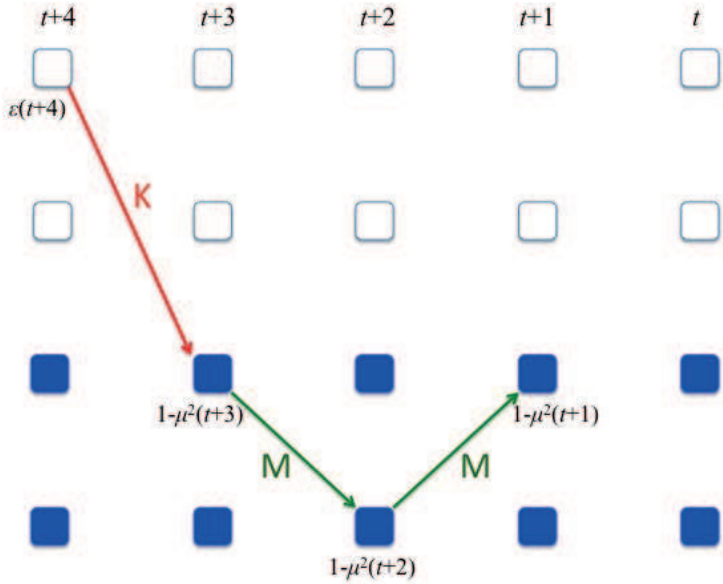}
\caption{Back-propagation of errors from the future through the hidden units.  The example path here starts at a visible unit $i$ where the output error $\epsilon_i(t+4)$ is measured.  It is then propagated back in time, first to a hidden unit at time $t+3$, then through another hidden unit at $t+2$ and finally to the one at $t+1$ which is the receiving unit on the connection being evaluated. It gives a change in that connection strength equal to the product of  $\epsilon_i(t+4)$, all the connection strengths on the path, and factors of $1-\mu_b^2(t)$ for each hidden unit on the path. The total connection strength change is a sum over all such paths from all visible units in the future.}\label{BPgraph}
 \end{center}
\end{figure}

This all has a nice graphical interpretation.  The effective error is the sum over all paths starting at future visible units (time $t+2+q$) and propagating back through the hidden units at intermediate times until it reaches the receiving unit $a$ at time $t+1$.  For each such path, we pick up a factor $\epsilon_i(t+q+2)$ at the visible error source, a factor of a $K_{ib}$ for backpropagating from the source unit to a hidden unit $b$, factors of elements of $\sf M$ for the hidden--to-hidden connections on the path, and factors of $X_{cc} = 1-\mu_c^2$ at every hidden unit $c$ that it passes through.  This is just the standard prescription for back-propagation of errors in layered networks.  Fig.~\ref{BPgraph} shows a typical path for $q =2$.

\subsection{Numerical results}

\begin{figure}[htp]
\begin{center}
	\subfigure{\includegraphics[width=2.4in]{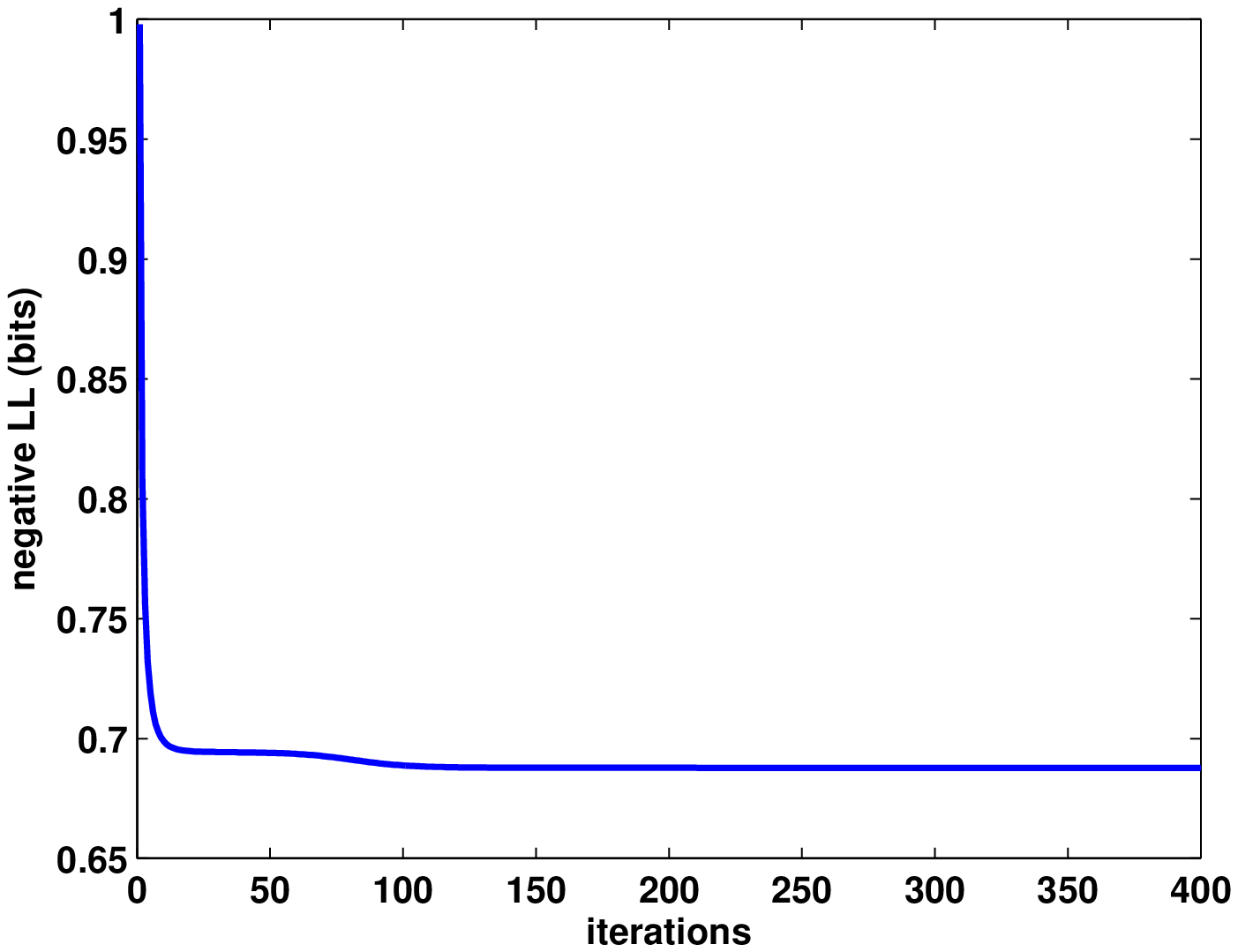}}
	\subfigure{\includegraphics[width=2.4in]{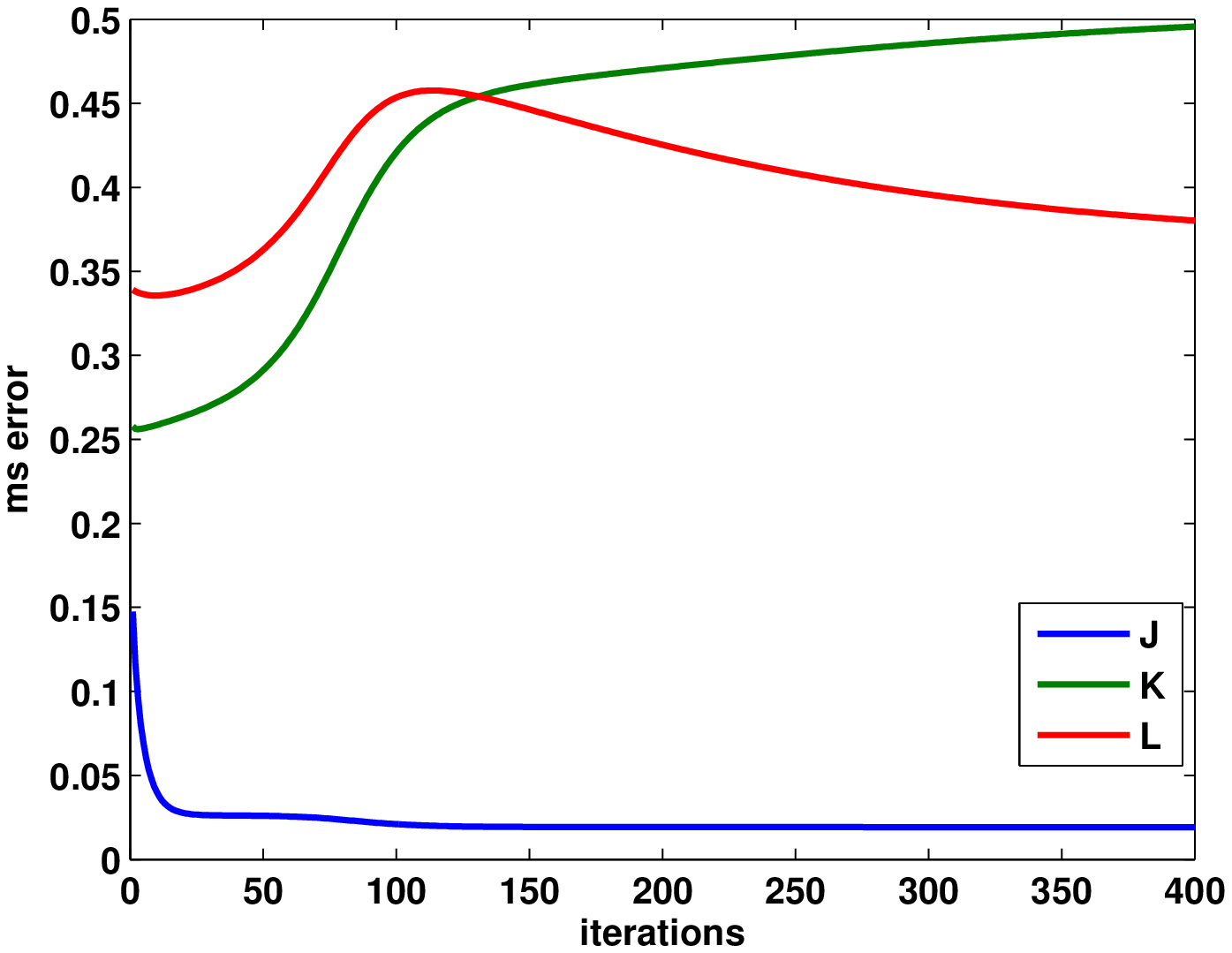}} \\
	\subfigure{\includegraphics[width=2.4in]{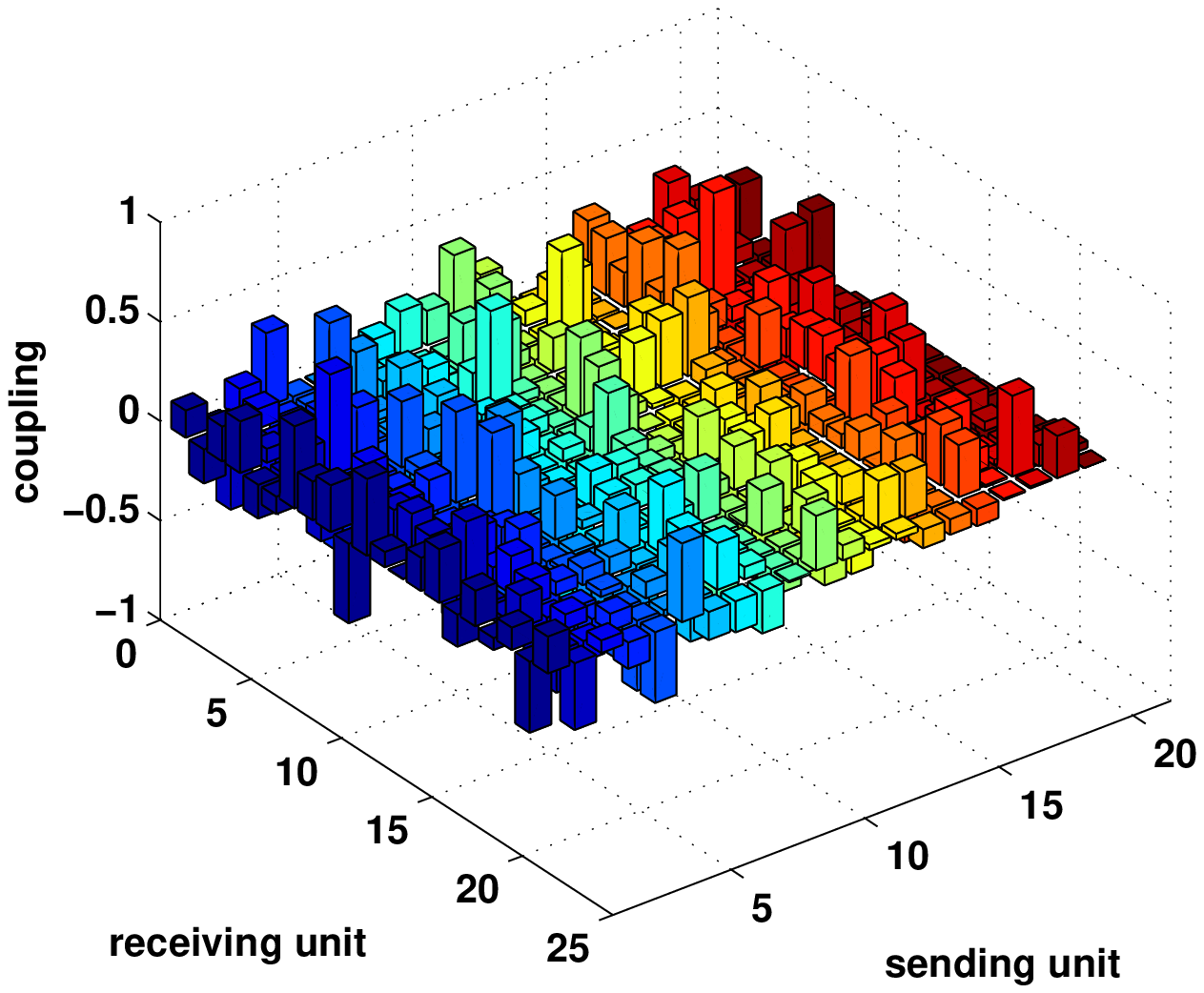}}
	\subfigure{\includegraphics[width=2.4in]{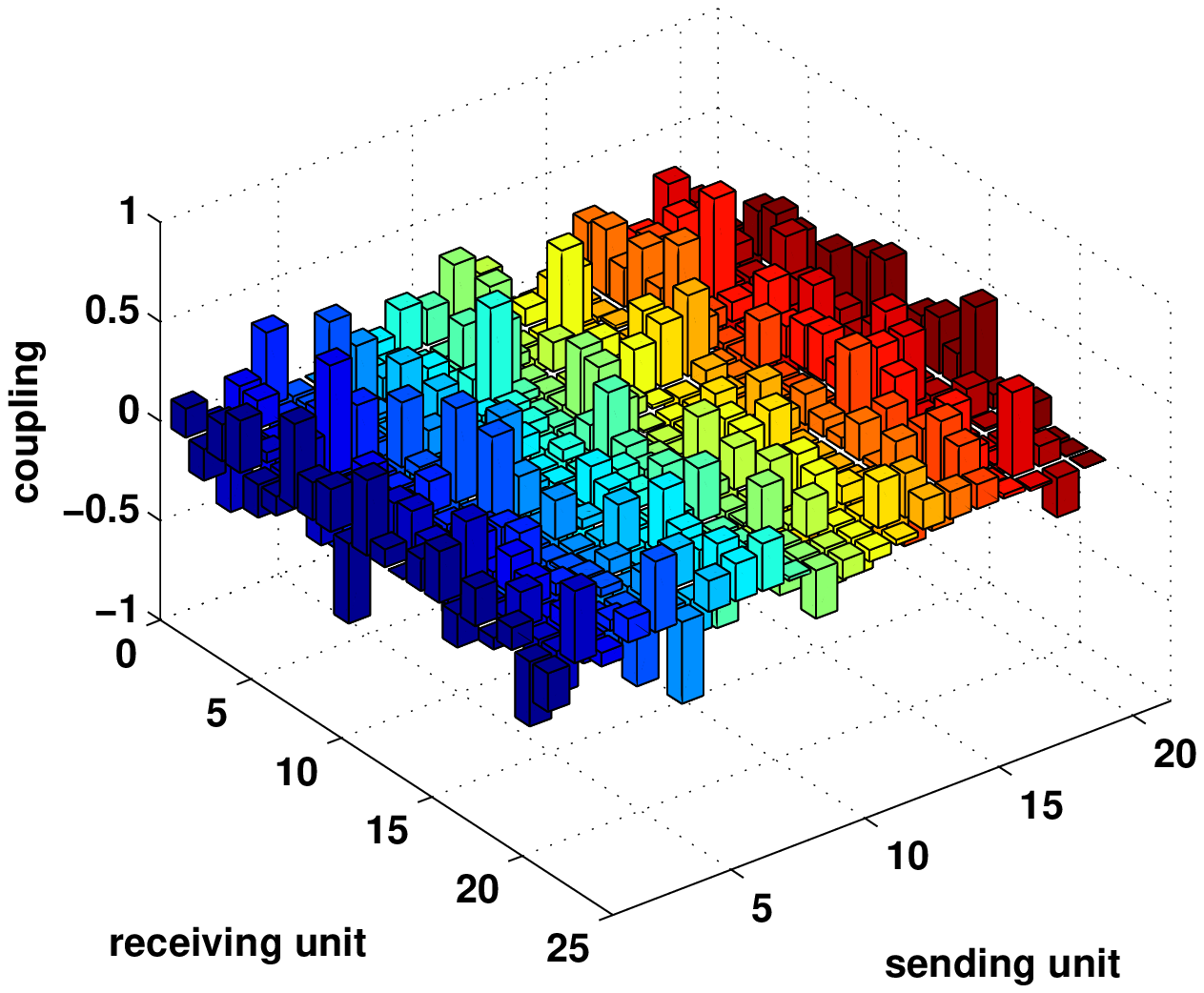}} \\
	\subfigure{\includegraphics[width=2.4in]{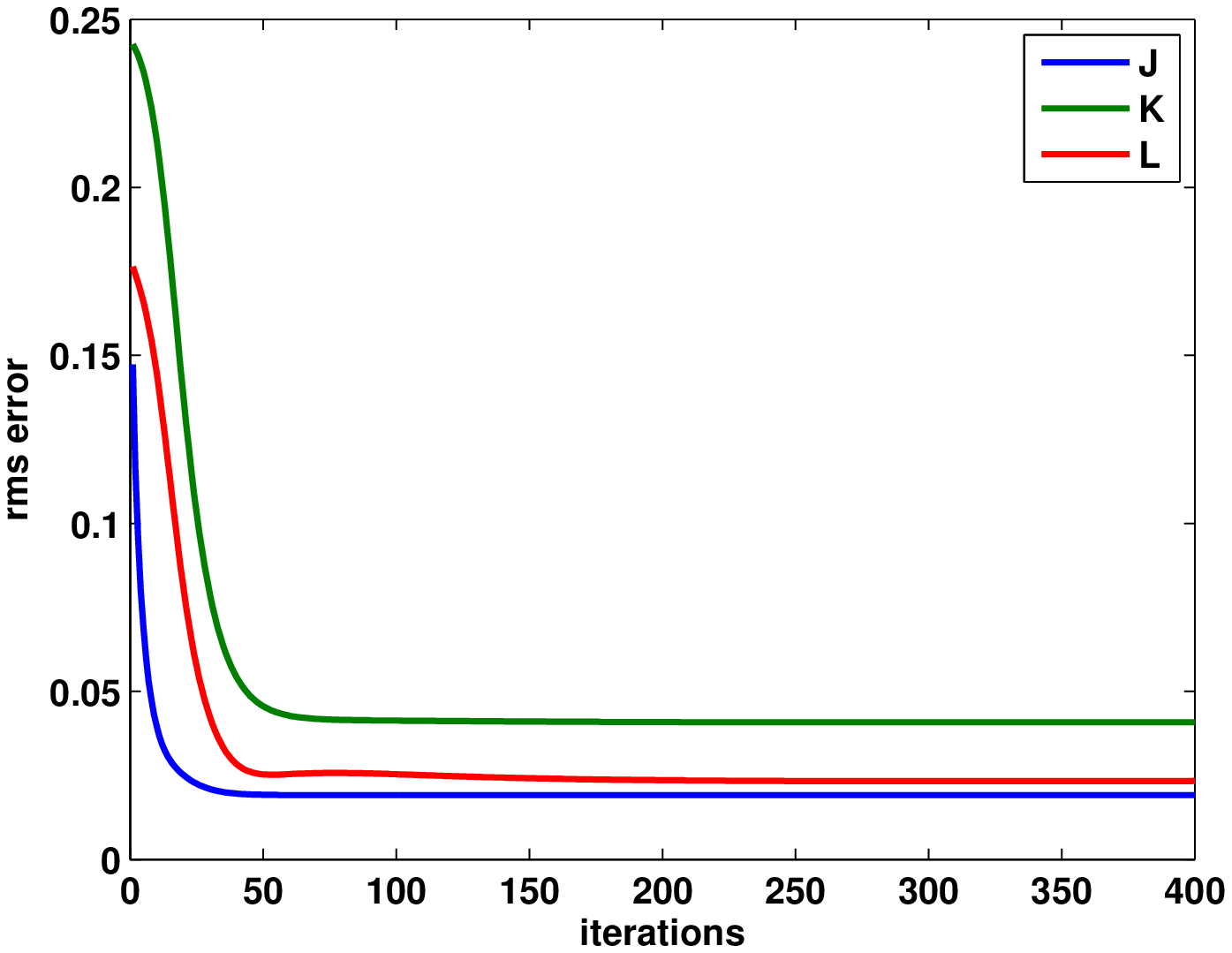}}
	\subfigure{\includegraphics[width=2.4in]{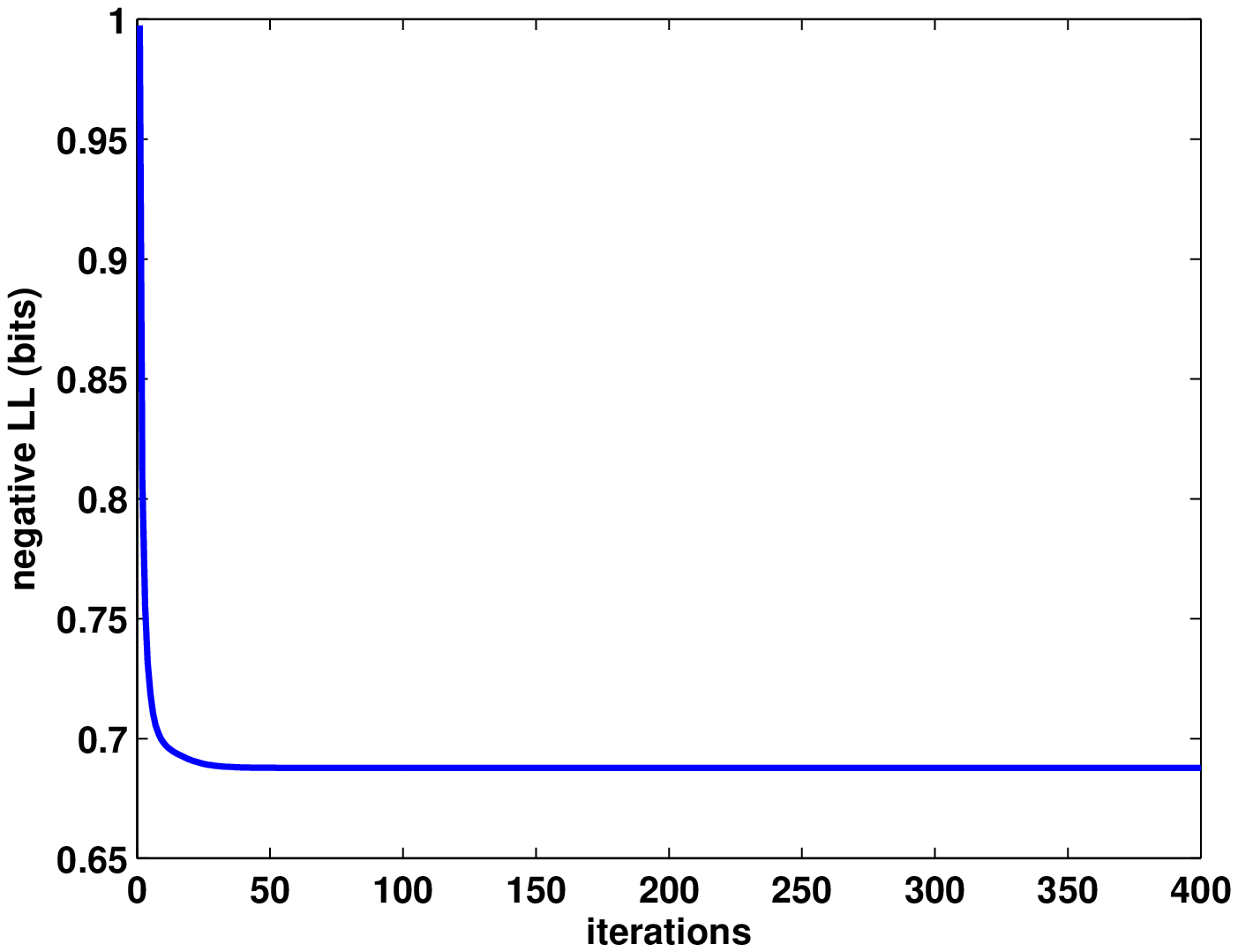}} 
\end{center}
\caption{Learning example: network of 18 visible and 2 hidden units (no hidden-hidden connections).  Top left: iterative minimization of the cost function $- {\mathcal L}$.  Top right: rms errors on $J_{ij}$, $K_{ib}$, and $L_{aj}$ as functions of the number of iterations of the learning algorithm  when it is started at small random values of the couplings.  Middle panels: true (left) and inferred coupling strengths. The hidden units are number 19 and number 20.  Bottom panels: rms errors (left) and cost function (right) when the initial parameter values have the correct signs. }     
\label{continhidlearning}
\end{figure}

In this calculations reported in this paper we restrict ourselves to networks with no hidden-to-hidden connections (${\sf M}=0$).  This simplifies the learning algorithm considerably: There are no backpropagation paths longer than two steps.  Fig. \ref{continhidlearning} shows an example of learning for a network with 18 visible and 2 hidden units, based on 10000 time steps of data.  The top left panel shows how the cost function (the negative log-likelihood of the data) falls smoothly to a minimum.  The top right panel shows the evolution of the errors in the couplings $J_{ij}$, $K_{ib}$ and $L_{aj}$ under learning.  The apparent poor performance can be understood by comparing the middle panels, which show the coupling matrix elements of the model that generated the data (left) and the inferred couplings (right), respectively.  It is apparent that the input connection strengths $L_{2j}$ to the second hidden unit (unit 20 in these plots) are negatives of each other in the two panels.   The same is true of the outgoing connections $K_{i2}$ from that unit, though it is hard to see in these graphs.  These two inversions have no effect on the visible units, so the ``true'' model and the one with the flipped signs of $L_{2j}$ and $K_{i2}$ are equivalent: there is no way we can know from the visible data alone which one was the true model.  The bottom left panel shows how, if we bias the initial random values of the couplings to have the right sign, results close to the true model are obtained.  The equivalence of the two inferred models is apparent from the fact (bottom right panel) that the final values of the cost function are exactly the same. 

In this example it was easy to see the relation between the inferred and true connections.  However, in general there is a $2^{N_h}\times N_h!$-fold degeneracy (the signs of the connections to and from every hidden unit could be flipped, and the labels on the hidden units can be permuted arbitrarily.)  Thus, for large $N_h$, most likely one will infer one of the models equivalent to the true one, but not the true one itself.

\section{Stochastic hidden units}

The case where all units in the model, including the hidden ones, are stochastic is more difficult, but it is the more interesting one from a theoretical point of view.  Denoting the hidden units by $\sigma_a(t)$, the dynamics are now given by
\begin{equation}
P[s_i(t+1),\sigma_a(t+1)] \{s(t),\sigma(t)\}] = \frac{\exp [s_i(t+1)H_i(t)]}{2 \cosh H_i(t)}	
\frac{\exp [\sigma_a(t+1)B_a(t)]}{2 \cosh B_a(t)}		\label{stochdyn}				\end{equation}
with
\begin{eqnarray}
H_i(t) &=& \sum_j J_{ij}s_j(t) + \sum_b K_{ib}\sigma_b(t)			\label{Hdefstochastic} \\
B_a(t) &= &\sum_j L_{aj}s_j(t) + \sum_b M_{ab} \sigma_b(t).		\label{Bdefstochastic}
\end{eqnarray}
We restrict our treatment to networks with weak dense random connections, $J_{ij}, L_{aj} = {\mathcal O}(1/\sqrt{N_v})$, $K_{ib}, M_{ab} = {\mathcal O}(1/\sqrt{N_h})$, so that $H_i(t)$ and $B_a(t)$ are of order $1$.

The likelihood of the history of the full system is 
\begin{equation}
P[s,\sigma] = \prod_{tia} P[s_i(t+1),\sigma_a(t+1)] \{s(t),\sigma(t)\}], 		\label{fulllikelihood}
\end{equation}
and the likelihood of the visible history is
\begin{equation}
P[s] = \sum_{\sigma} P[s,\sigma].								\label{vislikelihood}
\end{equation}
The distribution of the $\sigma$, conditional on the observed data, is
\begin{equation}
P[\sigma | s] = \frac{P[s,\sigma]}{P[s]}.							\label{cond}
\end{equation}
This has the form of a Gibbs distribution $Z_s^{-1} \exp (-E_s[\sigma])$, with
\begin{equation} 
E_s[\sigma] =  \log P[s,\sigma]									\label{energy}
\end{equation}
and
\begin{equation}
Z_s = P[s] = \sum_\sigma P[s,\sigma].							\label{Z}
\end{equation}
($Z_s$ also depends on all the model parameters $\{J_{ij},K_{ib},L_{aj},M_{ab}\}$, but to save some space we do not write that explicitly.)  To show the nature of the interactions in the energy $E_s[\sigma]$, we write it out explicitly:
\begin{eqnarray} 
E_s[\sigma] &=& - \sum_t \left\{ \sum_{ij}s_i(t+1)J_{ij}s_j(t) +  \sum_{ib}s_i(t+1)K_{ib}\sigma_b(t) 
\right.															\nonumber \\
&+& \sum_{aj}\sigma_a(t+1)L_{aj}s_j(t) + \sum_b \sigma_a(t+1)M_{ab}\sigma_b(t)  \nonumber \\
&-& \sum_i\log 2 \cosh \left[\sum_j J_{ij}s_j(t) + \sum_b K_{ib}\sigma_b(t)\right] 		\nonumber \\
&-& \left. \sum_a\log 2 \cosh \left[\sum_j L_{aj}s_j(t) + \sum_b M_{ab}\sigma_b(t) \right] \right\}.      \label{Efull}
\end{eqnarray}
The first term is just a constant (independent of the $\sigma$s), the next two are like external fields acting on the $\sigma_a(t)$ from the visible data $s_i(t \pm 1)$ one time step in the future and past, respectively, and the fourth term represents interactions between $\sigma$s at successive time steps. The final two terms are interactions among all the $\sigma$s at one time (but these terms do not couple $\sigma$s at different times).  Their non-polynomial form leads to important features in this problem that are not present in Boltzmann machines.

\subsection{Exact learning algorithm}

Just as for Boltzmann machines, we can derive an exact learning algorithm for the model parameters by gradient ascent on $Z_s$, the log likelihood of the visible history. It can be written
\begin{eqnarray}
 \Delta J_{ij} &\propto& \frac{\partial \log Z_s}{\partial J_{ij}}  
 = \sum_t [s_i(t+1) -\langle \tanh H_i(t)\rangle_{\sigma|s}]s_j(t)   			\label{dJ} \\
\Delta K_{ib} &\propto& \frac{\partial \log Z_s}{\partial K_{ib}}  
= \sum_t \langle [s_i(t+1) - \tanh H_i(t)]\sigma_b(t) \rangle_{\sigma|s}		\label{dK} \\
\Delta L_{aj} &\propto&  \frac{\partial \log Z_s}{\partial L_{aj}}
= \sum_t \langle \sigma_a(t+1) - \tanh B_a(t)\rangle_{\sigma|s}s_j(t)      		\label{dL} \\
\Delta M_{ab} &\propto&  \frac{\partial \log Z_s}{\partial M_{ab}} 
= \sum_t \langle [\sigma_b(t+1) - \tanh B_a(t)]\sigma_b(t)\rangle_{\sigma|s} 	\label{dM}
\end{eqnarray}
The averages $\langle \cdots \rangle_{\sigma|s}$ are over all hidden histories $\sigma(t)$, weighted by the probability $P[\sigma | s] = Z_s^{-1} \exp \{-E_s[\sigma]\}$  that they produce the known visible history $s$.  In each learning rule, the first term comes from differentiating the terms in the first two lines of (\ref{Efull}) and the second from differentiating one of the $\log 2 \cosh$ terms.

When there are no hidden-to-hidden connections $M_{ab}$, $P[\sigma|s]$ becomes a product of independent terms, one for each $t$.  The averages over $P[\sigma|s]$ in (\ref{dJ}-\ref{dM}) then involve sums over $2^{N_h}$ terms, where $N_h$ is the number of hidden units.  For small networks, they can be computed exactly in a reasonable time.  

\subsection{Numerical results}

\begin{figure}[htp]
\begin{center}
\includegraphics[width=5in]{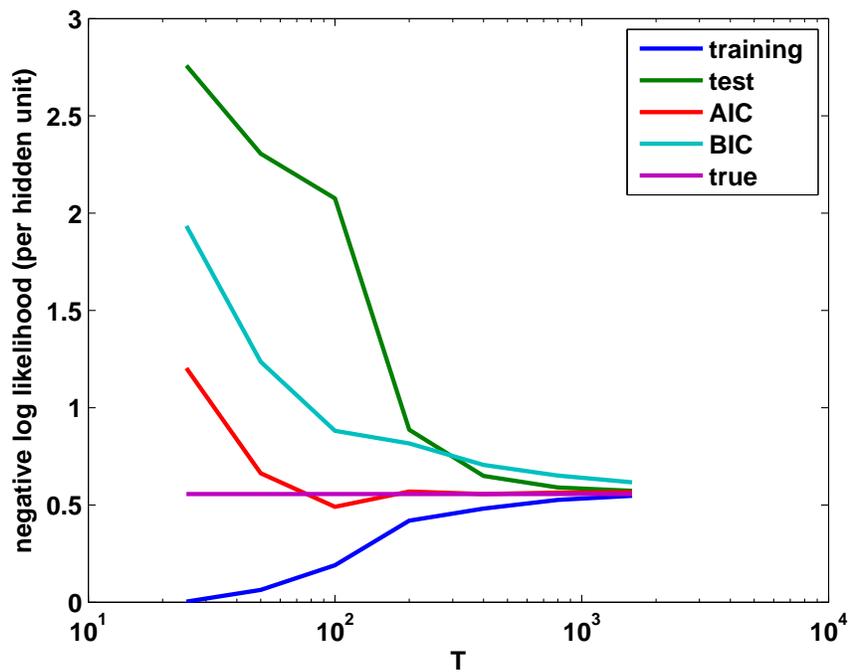}
\caption{Cost functions for learning in a network of 10 visible and 10 hidden units. (Color online) Blue: evaluated on training data. Green: evaluated on independent test data.  Red: Akaike information criterion (AIC \cite{AIC}).  Cyan: Bayesian information criterion (BIC \cite{BIC}).  Purple: $T \to \infty$ limiting value.}			\label{exactstochastic10}
\end{center}
\end{figure}

In Fig. \ref{exactstochastic10} we show, for a model with $N_h=N_v=10$ (and, again, no hidden-to-hidden connections), how the log-likelihood of the data converges to its asymptotic value as the number of steps in the data set is increased. All the couplings in this example were i.i.d. and normal with variance $0.1$.  In addition to the cost function $-\mathcal L$ evaluated on the training data, we also plot it evaluated on an independently-generated test data set.  We also plot the values of the Akaike and Bayesian information criteria, based on the training cost function.  The Akaike information criterion penalizes the estimated log likelihood (i.e., increases the cost) by the number of parameters $N$, and the Bayesian information criterion penalizes it by $N \log N$.    

For networks larger than $\sim 10$, one has to resort to Monte Carlo to estimate the averages.  When there are hidden-to-hidden connections, the number of states to sum over becomes $2^{N_hT}$, where $T$ is the number of time steps in the data.  In this case, exact calculations are never possible, even for just one hidden unit, and even Monte Carlo becomes impractical for moderate numbers of hidden units.

\section{Mean field theory for stochastic hidden units}

An attractive approximate alternative is mean field theory.  It can be formulated variationally \cite{Barberbook}: One seeks the best approximation to $P[\sigma|s]$ that factorizes over the different $\sigma_a(t)$.  Each such factor is parametrized by a single number: the probability that $\sigma_a(t) = +1$.  Equivalently (and conventionally), one can use the ``magnetization'', denoted $\mu_a(t)$, which is the difference between the probabilities to be $+1$ and $-1$.  The entire factorizable distribution is then parametrized by the set of magnetizations $\{\mu_a(t)\}$.  The learning proceeds in an EM fashion \cite{Rolf74,EMoriginal}, iterating the two steps: (1) For given coupling parameters, find the $\mu_a(t)$ that maximize the factorized $\log Z_s$, and (2), for these $\mu_a(t)$, improve the estimates of the coupling parameters as in rules (\ref{dJ}-\ref{dM}) but with the averages computed under the factorized approximate $P[\sigma|s]$.

\subsection{Derivation of mean-field theory}

Under the factorizability assumption, the likelihood of the visible data $\{s_i(t)\}$, given $\langle \sigma_a(t) \rangle = \mu_a(t)$, is
\begin{equation}
P_{MF}[\mu,s] = \exp \{S[\mu] - E_s[\mu] \} \equiv \exp A[\mu, \{J_{ij},K_{ib},L_{aj},M_{ab}\}]	,							\label{PMF}
\end{equation}
where
\begin{equation}
S[\mu] = -\sum_{at} \left[ \frac{1+\mu_a(t)}{2} \log \left ( \frac{1+\mu_a(t)}{2} \right ) +
 \frac{1-\mu_a(t)}{2} \log \left ( \frac{1-\mu_a(t)}{2} \right ) \right]     			\label{entropy}
\end{equation}
is the entropy: the average log of the probability of magnetizations $\mu_a(t)$.  
In (\ref{PMF}) we indicate explicitly that $A$ depends on the parameters $\{J_{ij},K_{ib},L_{aj},M_{ab}\}$ (through $E_s$).  Thus, the EM learning procedure involves repeatedly maximizing over $\mu$ for fixed parameters (the ``E-step'') and taking uphill steps on $A$ (equivalently, downhill steps on $E_s$) in parameter space for fixed $\mu$ (the ``M-step'').

The prescription for obtaining the average energy by the replacement $\sigma_a(t) \rightarrow \mu_a(t)$ in $E_s$ is based on the independence of different $\sigma_a(t)$ under the factorized distribution.  For example, if $E_s$ contains a term like $\sigma_a(t)\sigma_b(t)$, then $\langle \sigma_a(t)\sigma_b(t) = \langle \sigma_a(t)\rangle \langle \sigma_b(t)\rangle = \mu_a(t)\mu_b(t)$.  Thus, one might think that we get the $E_s[\mu]$ to use in (\ref{PMF}) by simply substituting $\mu$ for $\sigma$ in (\ref{Efull}).  Then maximizing $A[\mu]$ would lead to the equations
\begin{eqnarray}
&\tanh^{-1}\mu_a(t) = \sum_j L_{aj}s_j(t-1) +\sum_b M_{ab}\mu_b(t-1) 		\nonumber \\
&+ \sum_i \left\{ s_i(t+1) - \tanh  \left[\sum_j J_{ij}s_j(t) + \sum_b K_{ib}\mu_b(t)\right] \right\}K_{ia} \nonumber \\
&+  \sum_b \left\{ \mu_b(t+1) - \tanh  \left[\sum_j L_{bj}s_j(t) + \sum_c M_{bc} \mu_c(t)\right]	
\right\} M_{ba}   												\label{nMF}
\end{eqnarray}
for the $\mu_a(t)$.   This equation has a nice interpretation:  The first two terms are just the inputs from visible and hidden units, respectively, at the previous time step, and the last two terms are just the back-propagated errors from visible and hidden units one time step later.  

However appealing this equation looks, it is wrong.   One has to be careful in the $\log 2 \cosh$ terms in $E_s[\sigma]$.   Expanding it in powers of the $K_{ib}$, we get a second order term proportional to $\sum_{ab}K_{ia}K_{ib}\sigma_a\sigma_b$.  The double sum includes terms with $a=b$, and for these terms we should make the replacement $\sigma_a^2 = 1$, not $\sigma_a^2 = \mu_a^2$.  (This situation does not arise for the usual Ising energy $- \sum_{i < j} J_{ij} s_is_j$, since the $i=j$ term is explicitly excluded from the sum.)   The same problem comes up in all higher-order terms in the expansion whenever there are repeated indices in the sums over hidden unit indices.  This problem was noticed already by Saul {\em et al} \cite{Sauletal}, who tried to deal with it by introducing an extra set of variational parameters.  Here, we make a treatment for a particular ensemble of models that is exact (within the factorization approximation) in the limit of large $N_h$.  In these models, all the couplings are zero-mean independent random numbers with variances  
proportional to $1/N$.  This makes the net inputs $H_i(t)$ and $B_a(t)$ Gaussian (for large $N$), with variances of order unity.  

Writing the $n$th order term in the expansion of $\log 2 \cosh H$ (we drop the visible unit index $i$ temporarily here, for simplicity) as
\begin{equation}
\alpha_n = \frac{c_n}{n!}\sum_{a_1 \cdots a_{n}} 
K_{a_1} \cdots K_{a_{n}} \sigma_{a_1} \cdots \sigma_{a_{n}} ,      \label{nthorderterm}
\end{equation}
consider first the terms in every term of (\ref{nthorderterm}) where two of the indices are equal.  There are $n(n-1)/2$ such pairs, so the correction to this subset of the $n$th order terms is
\begin{equation}
\gamma_n^{(2)} = \half \frac{c_n}{(n-2)!}\sum_{a_1, a_2, \cdots a_{n-2}} 
 K_{a_1}  \cdots K_{a_{n-2}} \sigma_{a_1} \cdots \sigma_{a_{n-2}} \left[\sum_a K_a^2(1-\mu_a^2)\right].
\label{nthorderpairs}
\end{equation}
because naive substitution of $\mu_a$ for $\sigma_a$ would have given $\mu_a^2$ instead of $1$.  But what multiplies the sum on $a$ here is just half the $n-2$nd term in the expansion of the second derivative of $\log 2\cosh H$, i.e., $1 - \tanh^2 H$.  So we can sum all such terms over $n$, yielding a correction
\begin{equation}
E_2 = {\mbox{$\frac{1}{2}$}} (1 -\tanh^2 H)\sum_a K_a^2(1-\mu_a^2).						\label{E2}
\end{equation}
Thus, at this level of approximation, we should use an energy $E_s[\mu]$ in which $\sum_i \log 2 \cosh H_i(t)$ is replaced by
\begin{equation}
\sum_i \log 2 \cosh H_i(t) +  {\mbox{$\frac{1}{2}$}} \sum_{ia}[1 -\tanh^2 H_i(t)] K_{ia}^2 [1 - \mu_a^2(t)]                         \label{correctedEH}
\end{equation}
(now with the substitution $\sigma \rightarrow \mu$ in the first term). This looks like the TAP term in the free energy for the usual Ising model, but with the opposite sign.  The same argument applies to the $\log 2 \cosh B$ term in $E_s$, which should be replaced by 
\begin{equation}
\sum_a \log 2 \cosh B_a(t) +  {\mbox{$\frac{1}{2}$}} \sum_{ab}[1 -\tanh^2 B_a(t)] M_{ab}^2 [1 - \mu_b^2(t)]  .                       \label{correctedEB}
\end{equation}
These corrections will lead to new terms in the MF equations for $\mu_a(t)$ and in the learning rule for the $K_{ia}$ and $M_{ab}$.   Note also that, for the models we are considering, these correction terms are of order $1$ (per visible or hidden unit, respectively) since they contain sums of $N_h$ terms and each term is of order $1/N_h$.

We can also sum up terms with 2 pairs of indices equal, 3 pairs of indices, equal, etc.  Consider first the terms where two pairs of indices are equal.  In the $n$th order term $\alpha_n$ (\ref{nthorderterm}), there are $n!/[4! (n-4)!]$ ways of picking the 4 indices and 3 ways to pair them.  The correction is
\begin{equation}
\gamma_n^{(4)} = \frac{3}{4!} \frac{c_n}{(n-4)!} \sum_{a_1, a_3, \cdots a_{n-4}} K_{a_1}  \cdots K_{a_{n-4}}\sigma_{a_1}\cdots \sigma_{a_{n-4}}.\left[\sum_a K_a^2(1-\mu_a^2)\right]^2  \label{4th}
\end{equation}
The sum over $n$ of these terms is just 
\begin{equation}
E_4 = \frac{3}{4!}\frac{\partial^4 (\log 2 \cosh H)}{\partial H^4}\left[\sum_a K_a^2(1-\mu_a^2)\right]^2. 															\label{sum4th}
\end{equation}												
Like (\ref{E2}),  (\ref{4th}) is of order $1$.

Extending this argument to the general term with $j/2$ pairs of coincident indices,  in the $n$th order term, there are $n!/[j!(n-j)!]$ ways to pick our the $j$ indices, and the number of ways to pair them is $(j-1)!! \equiv (j-1)(j-3) \cdots 3 \cdot 1$.  Thus, we get a correction
\begin{equation}
E_j = \frac{(j-1)!!}{j!} \frac{\partial^j (\log 2 \cosh H)}{\partial H^j}
\left[\sum_a K_a^2(1-\mu_a^2)\right]^j.								\label{sumjth}
\end{equation}
Again, all these terms are all of order $1$.  

On the other hand, terms we have not considered, with $3$ or more indices equal, are negligible in the mean-field limit $N_h \to \infty$.  (Consider terms with $p$ equal indices.  They involve the sum $\sum_aK_{a}^p$, which is of order $N_h^{1-p/2}$ and therefore negligible for $p>2$ as $N_h \to \infty$. 

Now we can sum all the $E_j$ over $j$, exploiting the fact that $(j-1)!!$ is the $j$th moment of a zero-mean univariate normal distribution.  The result of all these manipulations is simply the replacement
\begin{equation}
\log 2 \cosh H_i(t) \longrightarrow
\int Dx \log 2 \cosh [H_i(t) + \Delta_i(t) x],								\label{result}
\end{equation}
where
$Dx$ means $(2\pi)^{-1/2}{\rm e}^{-x^2/2}dx$ and
\begin{equation}
\Delta_i^2(t) = \sum_a K_{ia}^2[1-\mu_a^2(t)].							\label{Deltaidef}
\end{equation}
Thus, the effect of all these corrections can be described in terms of an effective Gaussian noise.  The same arguments apply to the $\log 2 \cosh B$ term, with the final result that the effective energy can be written, exactly in the limit $N_h \to \infty$, as
\begin{eqnarray} 
E_s[\mu] &=& - \sum_t \left\{ \sum_{ij}s_i(t+1)J_{ij}s_j(t) +  \sum_{ib}s_i(t+1)K_{ib}\mu_b(t) 
\right.														\nonumber \\
&+& \sum_{aj}\mu_a(t+1)L_{aj}s_j(t) + \sum_b \mu_a(t+1)M_{ab}\mu_b(t)  \nonumber \\
&-& \sum_i \int Dx \log 2 \cosh \left[\sum_j J_{ij}s_j(t) + \sum_b K_{ib}\mu_b(t) + \Delta_i(t)x\right] 															\nonumber \\
&-& \sum_a \left. \int  Dy \log 2 \cosh \left[\sum_j L_{aj}s_j(t) + \sum_b M_{ab}\mu_b(t) + \Gamma_a(t)y\right] \right \}  ,      												\label{Eefffull}
\end{eqnarray}
with
\begin{equation}
\Gamma_a^2(t) = \sum_b M_{ab}^2[1-\mu_b^2(t)].					\label{Gammaadef}
\end{equation}

We note that this form could have been motivated heuristically:  In (\ref{Hdefstochastic}) and (\ref{Bdefstochastic}), the $\sigma_b(t)$ are fluctuating variables of variance $1-\mu_b^2(t)$.  Since $K_{ib}$ and $M_{ab}$ are assumed to be independent random variables, $H_i(t)$ and $B_a(t)$ are normally distributed with variances $\Delta_i^2(t)$ and $\Gamma_a^2(t)$ given by (\ref{Deltaidef}) and (\ref{Gammaadef}), respectively.

\subsection{Learning algorithm}

The resulting equations for the E-step are then
\begin{eqnarray}
&\tanh^{-1}\mu_a(t) = \sum_j L_{aj}s_j(t-1) +\sum_b M_{ab}\mu_b(t-1) 		\nonumber \\	
&+ \sum_i \left \{s_i(t+1) - \int Dx \tanh \left[\sum_j J_{ij}s_j(t) + \sum_b K_{ib}\mu_b(t) +\Delta_i(t)x\right]\right\}K_{ia} 		\nonumber \\
&+ \mu_a \sum_i \left\{1-\int Dx\tanh^2\left[\sum_j J_{ij}s_j(t) + \sum_b K_{ib}\mu_b(t)+\Delta_i(t)x\right]\right\}K_{ia}^2		\nonumber \\
&+ \sum_b\left\{\mu_b(t+1) - \int Dy \tanh \left[\sum_j L_{bj}s_j(t) + \sum_c M_{bc}\mu_c(t)  +\Gamma_b(t)y\right]\right \}M_{ba}    \nonumber \\
&+ \mu_a \sum_b \left\{1-\int Dy\tanh^2\left[\sum_j L_{bj}s_j(t) + \sum_c M_{bc}\mu_c(t) +\Gamma_b(t)y\right]\right\}M_{ba}^2.		\label{MFmu}
\end{eqnarray}
They differ from the naive equations (\ref{nMF}) in that the $\tanh$ terms in the second and fourth lines are averaged over the Gaussian noises and in the presence of the new terms on the third and fifth lines.  The latter have the form of cavity field corrections \cite{MPV}:  The effect of $\mu_a$ itself on the expected $s_i(t+1)$ and $\mu_b(t+1)$ should not be counted in calculating the $\tanh H$ terms in the second and fourth lines.

For the M-step, the learning rules for $J_{ij}$ and $L_{aj}$ are 
\begin{eqnarray}
 \Delta J_{ij} &\propto& -\frac{\partial E_s}{\partial J_{ij}}  
 = \sum_t \left\{ s_i(t+1) -\int Dx \tanh \left[H_i(t) + \Delta_i(t)x\right] \right\}s_j(t)   \label{dJMF} \\
\Delta L_{aj} &\propto&  -\frac{\partial E_s}{\partial L_{aj}}
= \sum_t \left\{ \mu_a(t+1) - \int Dy \tanh \left[B_a(t)+\Gamma_a(t)y\right]\right\}s_j(t) ,
															\label{dLMF} 
\end{eqnarray}  
differing from those we would find in the naive mean field theory only in the averaging of the $\tanh$'s over the Gaussian noises.  The rules for $K_{ib}$ and $M_{ab}$,
\begin{eqnarray}
\Delta K_{ib} &\propto& -\frac{\partial E_s}{\partial K_{ib}}  
 = \sum_t \left\{ \left( s_i(t+1) -\int Dx \tanh \left[H_i(t) + \Delta_i(t)x\right] \right)\mu_b(t) \right.   \nonumber \\
 &-&  \left.  \left[1-\int Dx\tanh^2 [H_i(t) + \Delta_i(t)x] \right] K_{ib} [1-\mu_b^2(t)] \right\} 
 															    \label{dKMF} \\
\Delta M_{ab} &\propto&  -\frac{\partial E_s}{\partial M_{ab}}
= \sum_t \left\{ \left(\mu_a(t+1) - \int Dy \tanh \left[B_a(t)+\Gamma_a(t)y\right]\right)\mu_b(t)\right.
\nonumber \\
&-&  \left. \left[1-\int Dy \tanh^2 [B_a(t) + \Gamma_a(t)y] \right] M_{ab} [1-\mu_b^2(t)] \right\} 
															\label{dMMF} 
\end{eqnarray}  
have extra terms that come from the dependence of $\Delta_i(t)$ and $\Gamma_a(t)$ on $K_{ia}$ and $M_{ab}$ in (\ref{Deltaidef}) and (\ref{Gammaadef}), respectively.  

For small $K_{ia}$ and $M_{ab}$ (i.e., at the level of the corrections (\ref{correctedEH}) and (\ref{correctedEB}), the E-step equations reduce to
\begin{eqnarray}
\tanh^{-1}\mu_a(t) &=& \sum_j L_{aj}s_j(t-1) +\sum_b M_{ab}\mu_b(t-1) 		\nonumber \\	
&+& \sum_i \left \{[s_i(t+1) - \tanh H_i(t)] K_{ia} + [1-\tanh^2 H_i(t)]K_{ia}^2\mu_a(t)	\right.	\nonumber \\
&+& \left.  \tanh H_i(t) [1-\tanh^2 H_i(t)]K_{ia}\sum_bK_{ib}^2[1-\mu_b^2(t)] \right\}   \nonumber \\
&+& \sum_b\left\{ [\mu_b(t+1) - \tanh B_b(t) ]M_{ba}    +[1-\tanh^2B_b(t)]M_{ba}^2\mu_a(t) \right.\nonumber \\
&+&\left. \tanh B_b(t) [1-\tanh^2 B_b(t)]M_{ba}\sum_cM_{bc}^2[1-\mu_c^2(t)] \right\}, 	\label{MFmuT}
\end{eqnarray}
and the learning rules are
\begin{eqnarray}
\Delta J_{ij} &\propto& \sum_t \left\{ s_i(t+1) - \tanh H_i(t)[1-(1-\tanh^2 H_i(t))\Delta_i(t) ]   \right\}s_j(t)   \label{dJMFT} 	\\							
\Delta L_{aj} &\propto&  \sum_t \left\{ \mu_a(t+1) - \tanh B_a(t)[1 -  (1-\tanh^2 B_a(t))\Gamma_a(t) ] \right\}s_j(t) , 													\label{dLMFT} \\
\Delta K_{ib} &\propto&   \sum_t \left\{ \left(s_i(t+1) -\tanh H_i(t)[1-(1-\tanh^2 H_i(t))\Delta_i(t) ] \right) \mu_b(t)     \right. \nonumber \\
 &-&  \left.  \left[1-\tanh^2 H_i(t)  \right] K_{ib} [1-\mu_b^2(t)] \right\} 
 \label{dKMFT} \\
\Delta M_{ab} &\propto&    
\sum_t \left\{ \left(\mu_a(t+1) -  \tanh B_a(t)[1 - (1-\tanh^2 B_a(t)) \Gamma_a(t) ] \right) \mu_b(t)\right.
\nonumber \\
&-&  \left. [1- \tanh^2 B_a(t)] M_{ab} [1-\mu_b^2(t)]  \right\} 
															\label{dMMFT} 
\end{eqnarray}  

A few final remarks are in order.  The reader might notice that the lowest-order corrections in (\ref{correctedEH}), (\ref{correctedEB}), and (\ref{MFmuT}) resemble Thouless-Anderson-Palmer (TAP) corrections in spin glasses \cite{TAP}. However, there the TAP equations come from the first corrections to the factorized-distribution approximation, whereas ours here come from evaluating the average energy within that approximation.  We expect that for our model here, as for spin glasses, to get an exact theory for large $N_h$, TAP corrections analogous to theirs should also be included.  We do not try to do that here, working entirely within the factorized-distribution ansatz.  In problems like ours for networks without hidden units, this is sometimes called ``naive mean field theory" \cite{RHPRL11}.

\subsection{Numerical results}

\begin{figure}[htp]
\begin{center}
	\subfigure{\includegraphics[width=2.4in]{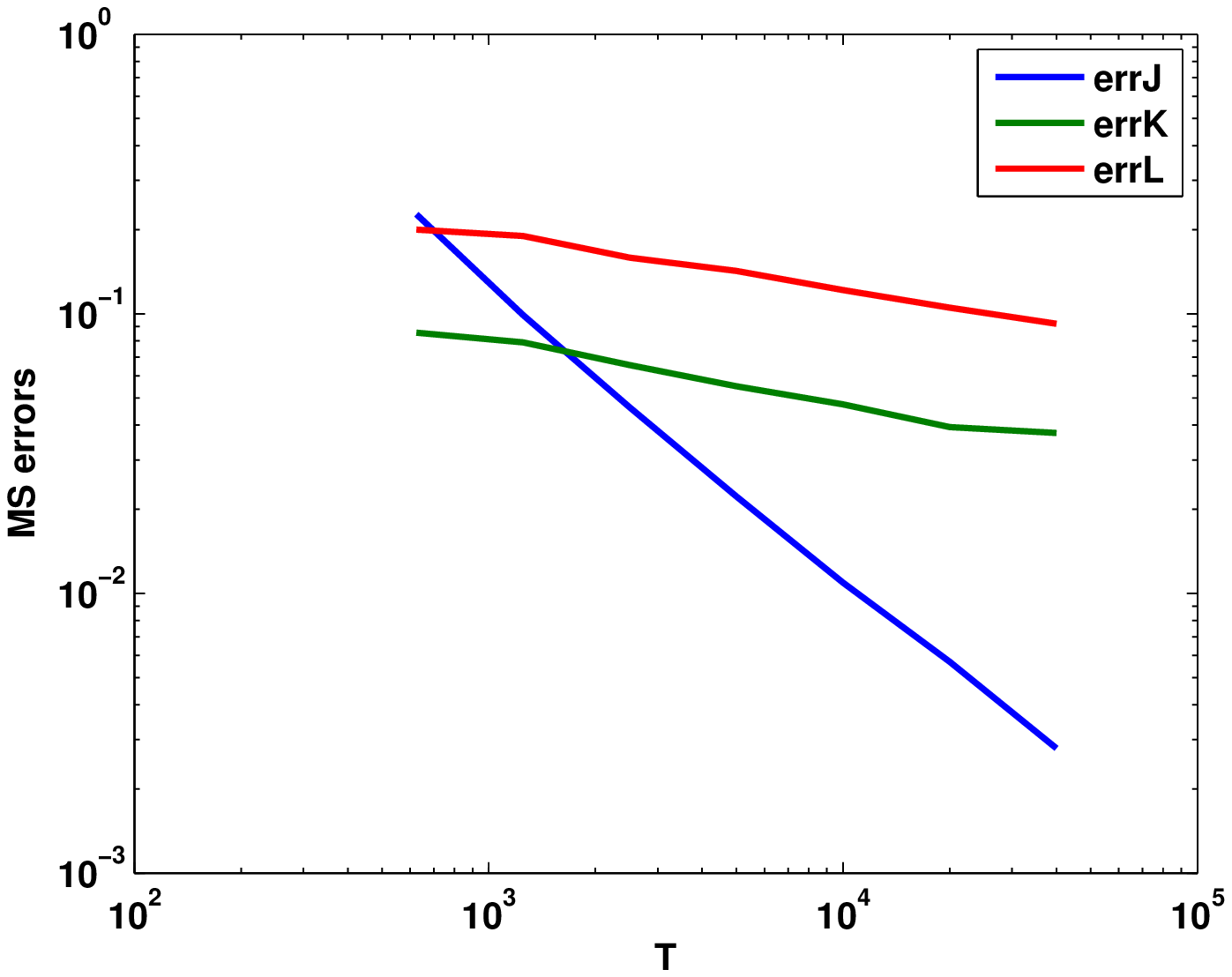}}
	\subfigure{\includegraphics[width=2.4in]{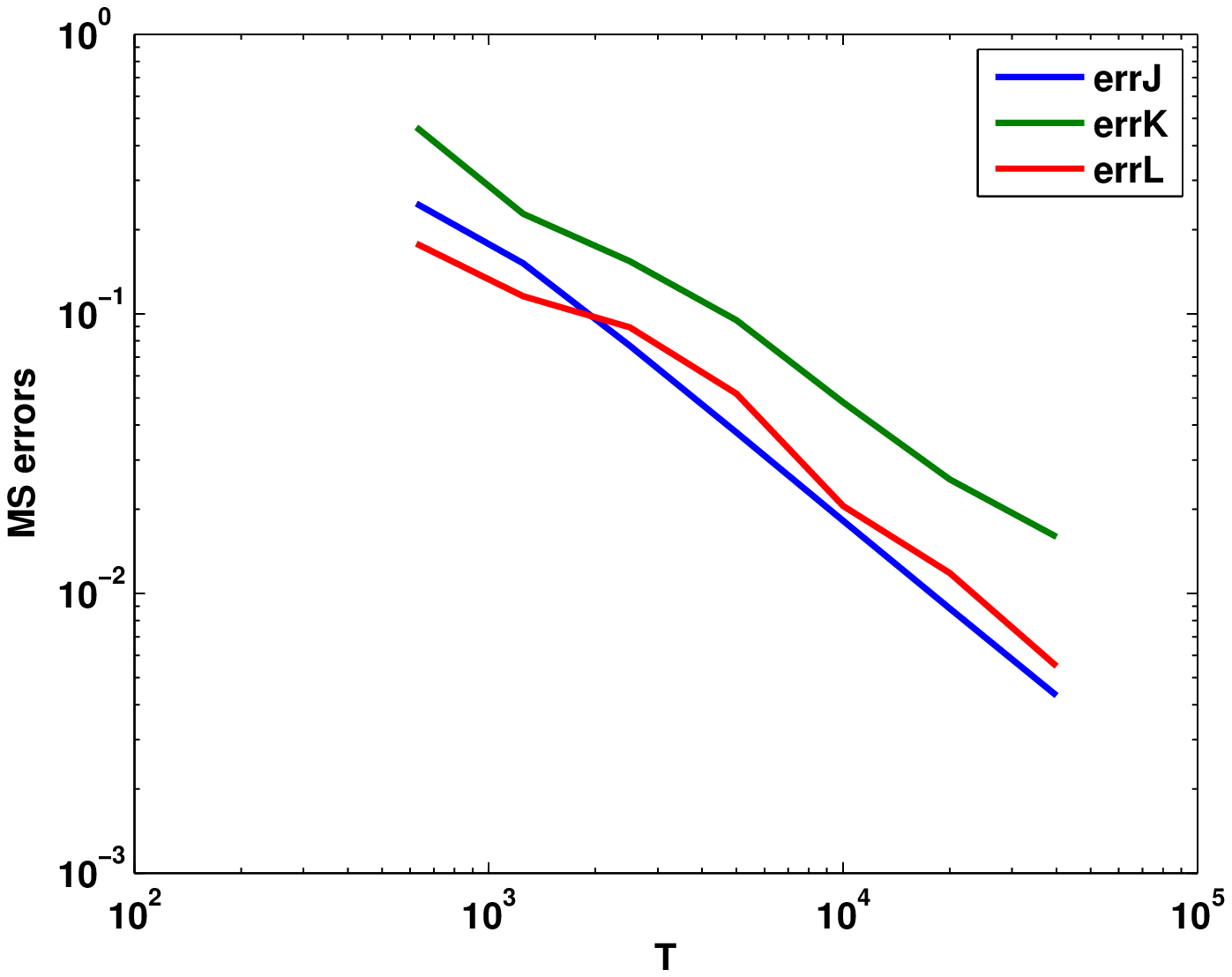}}  
\end{center}
\caption{Mean square errors on $J$s (blue), $K$s (green) and $L$s (red) computed in mean field theory as functions of data set length.  (Color online) Left panel: $N_v=N_h=80$. Right panel: $Nv=80$, $N_h=20$.  All couplings are i.i.d. normal, with variance $1/N_v$ for $J_{ij}$ and $L_{aj}$ and $1/N_h$ for $K_{ib}$. }     
\label{N80MF}
\end{figure}

We have carried out mean-field inference computations  for some models with no hidden-to-hidden connections ($M_{ab} = 0$), using the lowest-order mean-field equations (\ref{MFmuT}-\ref{dMMFT}).   Fig. \ref{N80MF} shows how the mean square errors of  $J_{ij}$, $K_{ib}$ and $L_{aj}$ depend on the data set length $T$ for two networks with $80$ visible units.  The left-hand panel shows the case where the number of hidden units $N_h = 80$, and the right-had panel shows the case where $N_h = 20$.  For the smaller $N_h$, all three mean square errors fall off like $1/T$, as we would expect to find if we could do this calculation exactly. However, for the larger $N_h$, while the errors on the visible-to-visible couplings also fall off with $T$ in this way, the errors on the couplings to and from the hidden units are larger and fall off much more slowly.

\begin{figure}[htp]
\begin{center}
	\subfigure{\includegraphics[width=2.4in]{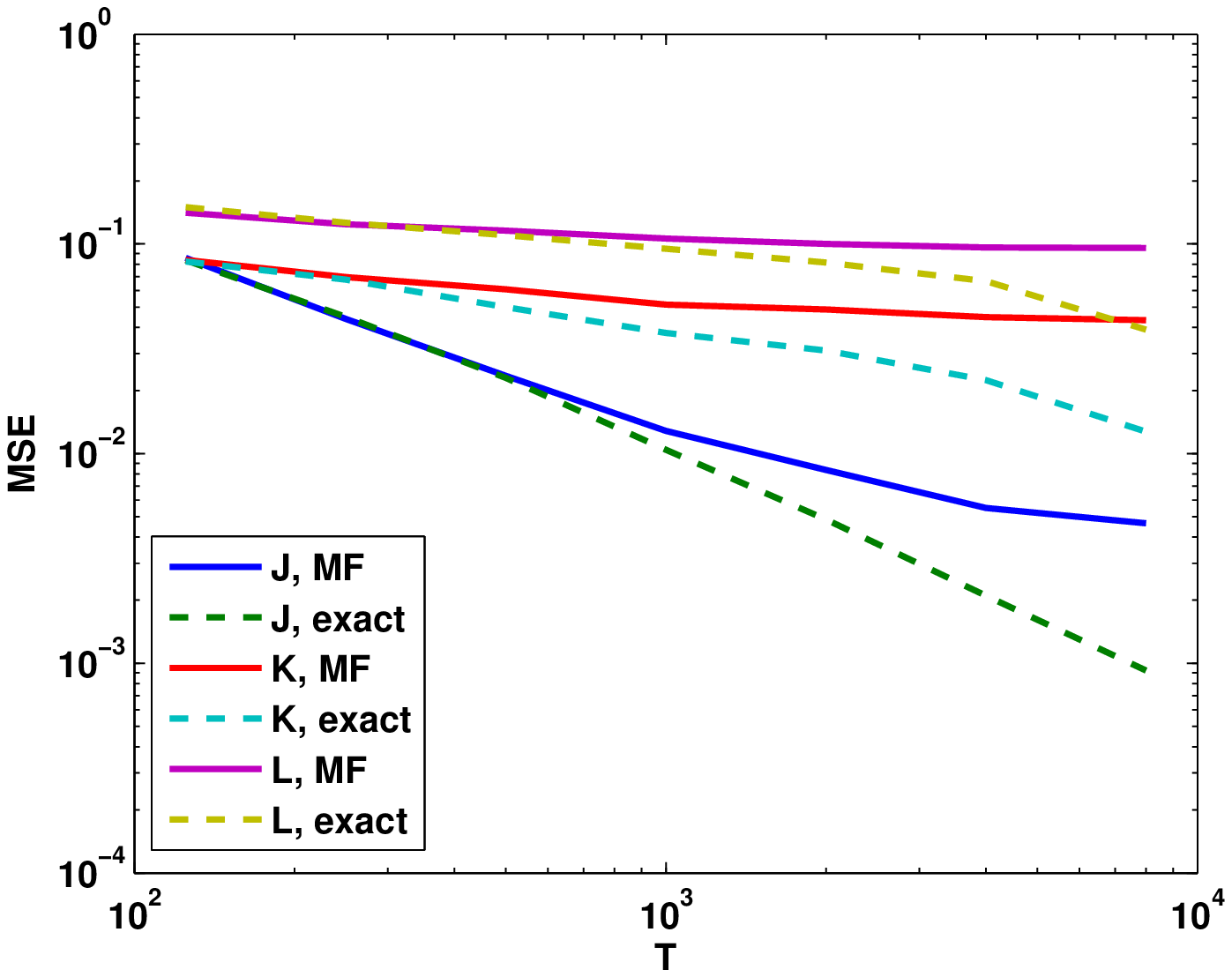}}
	\subfigure{\includegraphics[width=2.4in]{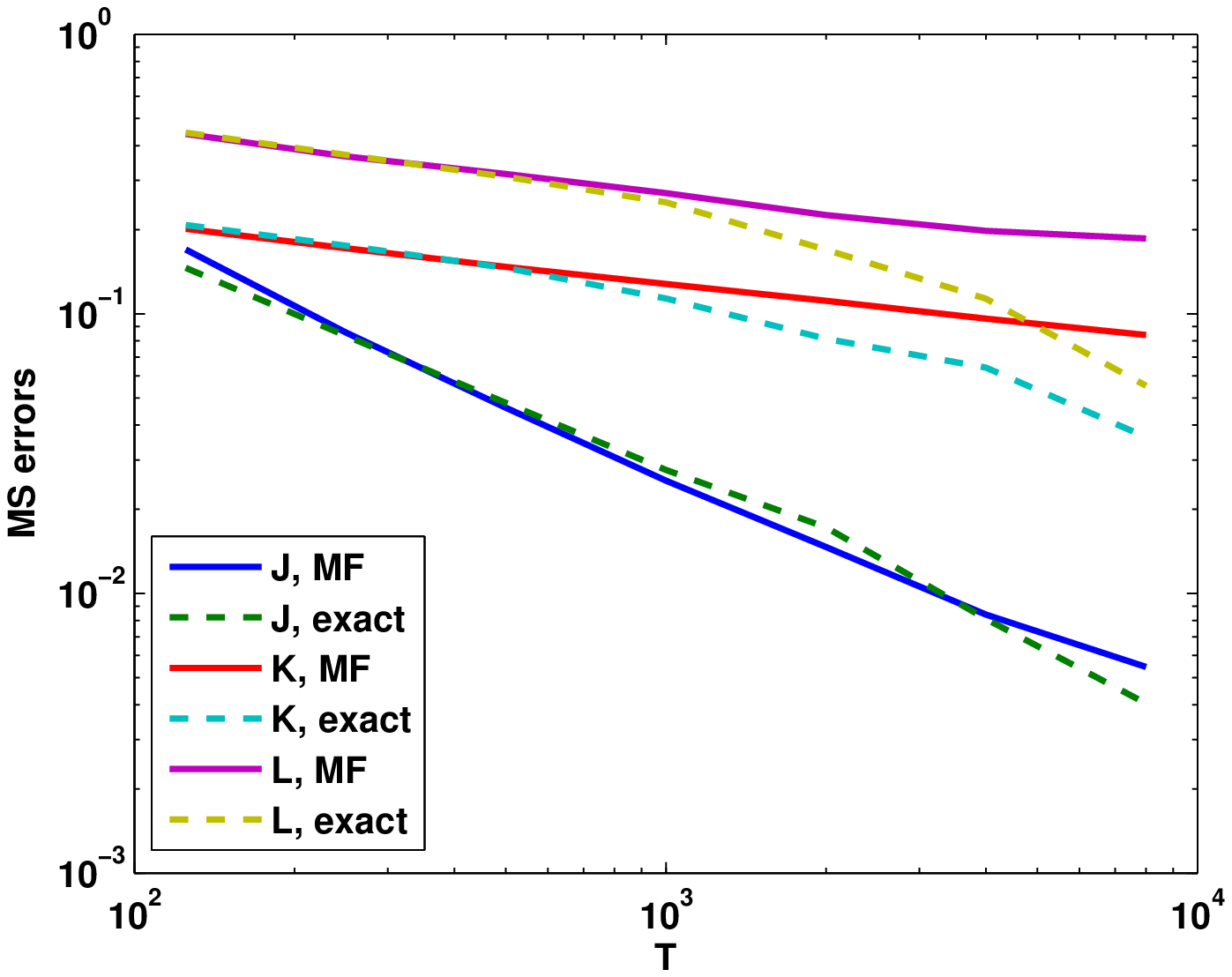}}  
\end{center}
\caption{Mean square errors on $J$s (blue), $K$s (green) and $L$s (red) as functions of data set length for small networks.  (Color online) Left Panel: $N_v=N_h=5$.  Right Panel: $N_v=N_h=8$.  Mean-field results are solid lines; exact results are dashed.  Couplings chosen as in Fig. 5.}     
\label{N5and8}
\end{figure}

We can get a little insight into this behavior by doing the mean-field calculations for small $N_h$, where it is also possible to do the exact calculations, as described in Sect. 2.3. Fig. \ref{N5and8} shows the results of both kinds of calculations for $N_v=N_h = 5$ and $8$.   In these cases we can see that for small $T$ the mean-field and exact calculations nearly coincide.  The $T$-dependence is in this region is qualitatively like that for the mean-field results at $N_v=N_h=80$.  However, at larger $T$, the mean-field errors all fall less rapidly.  At the same time, the exact calculation gives errors on the $J$s which continue to fall off like $1/T$, and those on the $K$s and $L$s also start to fall more rapidly at the largest $T$s studied.   This behavior is consistent with the expectation that, as for models with no hidden units, all exact-method errors should fall off asymptotically like $1/T$, while the mean-field errors should approach limits $\propto 1/N_h$ \cite{RHPRL11}.  However, apparently one has to go to very large data sets (roughly $T > 10^3N_h$) to see this.  

\section{Discussion}

We have derived learning rules for two kinds of stochastic binary networks with hidden units.  These networks differ from Boltzmann machines in that (1) the units in them are updated synchronously rather than asynchronously, and (2) the connection strengths are allowed to be asymmetric.   Because of these differences, the usual kind of Gibbs equilibrium does not hold, and a new kind of treatment is required.

The first kind of network has deterministic, continuous-valued hidden units.  The learning rules for it are very similar to those in the back-propagation-in-time approach for recurrent networks where all the units are deterministic and continuous-valued.  

Units in the second kind of network are binary and stochastic, like the visible units.  Here the learning problem is harder, but we have showed that one can always put it into the form of an equilibrium statistical mechanical problem with a non-polynomial energy function.  The learning rules involve averages over the Gibbs distribution for this problem.  For small systems and in the absence of hidden-to-hidden couplings, the problem can be solved exactly numerically, but otherwise one must resort to Monte Carlo methods or other approximations.  We explored in detail one such approximation: mean field theory.  A careful analysis revealed that the naive way one might write this theory was wrong, but we were able to construct a version of mean field theory that was exact for weak, dense connectivity in the limit of a large number of hidden units (the analog of the Sherrington-Kirkpatrick model of spin glasses \cite{SK}).  

We also performed some numerical calculations to illustrate and to begin to explore some of the features of the different kinds of networks and learning rules.   A general feature is that when the number of hidden units is large (i.e., comparable to the number of visible units), the errors in determining the couplings to and from the hidden units are much larger than those on the couplings among the visible units.  This is true for both kinds of networks and for both exact learning algorithms and mean field theory.  This should not be surprising, since the information about the connections to and from hidden units is only available indirectly, through the statistics of the visible units.  On the other hand, it is noteworthy that even a rather poor estimation of the connections to and from the hidden units does not spoil the good estimation of the couplings among the visible ones.  

Another point worth mentioning is that for small data lengths mean field theory is as good as doing the full exact calculation, which would take prohibitively long for $N_h$ much bigger than $10$ or so.  For large $N_h$ the errors on connections to and from hidden units can be rather large and fall off very slowly with $T$, but the results on small systems seem to show that doing the exact calculation instead of mean field theory (even if this were feasible), would not help except at very large $T$.     

We have only scratched the surface of this problem in our numerical calculations.  It would be useful to know,  for example, what the asymptotic errors on the $K$s and $L$s are for the mean-field algorithm in the limit of large data sets and at what $T$ the approach to these values begins, as functions of $N_h$ and $N_v$.   We leave this and other questions to future work.  The theory presented here provides a foundation for those investigations and, we hope, will point the way toward other questions that will be interesting to study.

\section*{Acknowledgements} We would like to thank Yasser Roudi and Ben Dunn, who have derived similar results for mean field theory in a different way, for discussions.


\begin{thebibliography}{99}

\bibitem{Schneidman}
     \newblock E. Schneidman, M. J. Berry, R. Segev, and W. Bialek, 
     \newblock \emph{Weak pairwise correlations imply strongly correlated network states in a neural population},
     \newblock Nature, \textbf{440} (2006), 1007--1012.
     
\bibitem{RTHPRE09}
     \newblock Y. Roudi, J. Tyrcha, and J. Hertz, 
     \newblock \emph{The Ising model for neural data: model quality and approximate methods for extracting functional connectivity},
     \newblock Phys Rev E, \textbf{79} (2009), 051915.
     
\bibitem{RHPRL11}
     \newblock  Y. Roudi and J. Hertz, 
     \newblock \emph{Mean-field theory for nonequilibrium network reconstruction},
     \newblock Phys Rev Lett, \textbf{106} (2011), 1048702.
     
\bibitem{chapter}
    \newblock J. Hertz, Y Roudi, and J Tyrcha,
    \newblock  \emph{Ising models for inferring network structure from spike data},
    \newblock  in ``Principles of Neural Coding" (eds. S. Panzeri and R. R. Quiroga),
                CRC Taylor and Francis, (2013), 527-546.     
    
\bibitem{AHS85}
     \newblock D. Ackley, G. E. Hinton, and T. J. Sejnowski, 
     \newblock \emph{A learning algorithm for Boltzmann machines},
     \newblock Cogn Sci, \textbf{9} (1985), 147--169.
     
\bibitem{Glauber}
     \newblock R. J. Glauber, 
     \newblock \emph{Time-dependent statistics of the Ising model},
     \newblock J Math Phys, \textbf{4} (1963), 294-307.
     
\bibitem{Peretto}
     \newblock P. Peretto, 
     \newblock \emph{Collective properties of neural networks: a statistical physics approach},
     \newblock Biol Cybern, \textbf{50} (1984), 51-62.     
     
\bibitem{RHW}
    \newblock D. E. Rumelhart, G. E. Hinton, and R. J. Williams,
    \newblock  \emph{Learning Internal Representations by Error Propagation},
    \newblock  vol. 1  chap. 8 in ``Parallel Distributed Processing" (eds. D. E. Rumelhart and J. L. McClelland), MIT Press, (1986).     
     
\bibitem{Pineda}
     \newblock  F. J. Pineda, 
     \newblock \emph{Generalization of back-propagation to recurrent neural networks},
     \newblock Phys Rev Lett, \textbf{59} (1987), 2229-2232.     

\bibitem{Pearlmutter}
     \newblock  B. A. Pearlmutter, 
     \newblock \emph{Learning state space trajectories in recurrent neural networks},
     \newblock Neural Computation, \textbf{1} (1989), 263-269.     
     
\bibitem{WilliamsZipser}
     \newblock  R. J. Williams and D Zipser, 
     \newblock \emph{A learning algorithm for continually running fully recurrent networks},
     \newblock Neural Comp, \textbf{1} (1989), 270-280.    
     
\bibitem{AIC}
     \newblock   H. Akaike, 
     \newblock \emph{A new look at the statistical model identification},
     \newblock IEE Transactions on Automatic Control, \textbf{19} (1974), 716-723.     

\bibitem{BIC}
     \newblock  G. E. Schwarz, 
     \newblock \emph{Estimating the dimension of a model},
     \newblock Annals of Statistics, \textbf{6} (1978), 461-464.     
      
     
\bibitem{Barberbook}
     \newblock D. Barber,
     \newblock ``Bayesian Reasoning and Machine Learning," chap. 11,
     \newblock Cambridge Univ. Press, 2012.
     
\bibitem{Rolf74}
     \newblock R. Sundberg, 
     \newblock \emph{Maximum likelihood theory for incomplete data from an exponential family},
     \newblock  Scand. J. Statistics, \textbf{1} (1974), 49-58.               
     
\bibitem{EMoriginal}
     \newblock A. P. Dempster, N. M. Laird, and D. B. Rubin, 
     \newblock \emph{Maximum likelihood from incomplete data via the EM algorithm},
     \newblock J. Roy. Stat. Soc. B, \textbf{39} (1977), 1-38.      
     
\bibitem{Sauletal}
     \newblock  L. K. Saul, T. Jaakkola, and M. I. Jordan, 
     \newblock \emph{Mean field theory for sigmoid belief networks},
     \newblock J. Art. Intel. Res, \textbf{4} (1996), 61-76.              
     
\bibitem{MPV}
      \newblock M. M{\'e}zard, G. Parisi, and M. Virasoro,
     \newblock ``Spin Glass Theory and Beyond," chap. 2,
     \newblock World Scientific, 1987.
     
\bibitem{TAP}
     \newblock D. J. Thouless, P. W. Anderson, and R. G. Palmer, 
     \newblock \emph{Solution of ``soluble model of a spin glass''},
     \newblock  Philos. Mag., \textbf{92} (1974), 272-279.               

\bibitem{SK}
     \newblock  D. Sherrington and S. Kirkpatrick, 
     \newblock \emph{Solvable model of a spin glass},
     \newblock Phys Rev Lett, \textbf{35} (1975), 1792-1796.
 
\end{thebibliography}
\end{document}